\documentclass[twocolumn,showpacs,preprintnumbers,amsmath,amssymb]{revtex4}
\usepackage{graphicx}% Include figure files
\usepackage{dcolumn}% Align table columns on decimal point
\usepackage{bm}% bold math

\begin{document}

\preprint{APS/123-QED}

\title{Temperature dependence of antiferromagnetic susceptibility in ferritin}

\author{N. J. O. Silva}
\thanks{On leave from the Departamento de F\'{\i}sica and CICECO, Universidade de Aveiro, Portugal}
\email{nunojoao@unizar.es}

\author{A. Mill\'{a}n}
\author{F. Palacio}
\affiliation{Instituto de Ciencia de Materiales de Arag\'{o}n,
CSIC - Universidad de Zaragoza. Departamento de Fisica de la
Materia Condensada, Facultad de Ciencias, 50009 Zaragoza, Spain.}

\author{E. Kampert}
\author{U. Zeitler}
\affiliation{Radboud University Nijmegen, High Field Magnet
Laboratory , NL-6525 ED Nijmegen, Netherlands}

\author{H. Rakoto}
\affiliation{LNCMP, 143 Avenue de Rangueil F-31400 Toulouse,
France}

\author{V. S. Amaral}

 \affiliation{Departamento de F\'{\i}sica and CICECO, Universidade de Aveiro, 3810-193 Aveiro, Portugal}

\date{\today}% It is always \today, today,

\begin{abstract}
We show that antiferromagnetic susceptibility in ferritin
increases with temperature between 4.2 K and 180 K (i. e. below
the N\'{e}el temperature) when taken as the derivative of the
magnetization at high fields ($30\times10^4$ Oe). This behavior
contrasts with the decrease in temperature previously found, where
the susceptibility was determined at lower fields ($5\times10^4$
Oe). At high fields (up to $50 \times10^4$ Oe) the temperature
dependence of the antiferromagnetic susceptibility in ferritin
nanoparticles approaches the normal behavior of bulk
antiferromagnets and nanoparticles considering
superantiferromagnetism, this latter leading to a better agreement
at high field and low temperature. The contrast with the previous
results is due to the insufficient field range used ($< 5
\times10^4$ Oe), not enough to saturate the ferritin uncompensated
moment.
\end{abstract}

% saturation and magnetic susceptibility, Antiferromagnetism, magnetization curves, fine particles (respectively)
\maketitle

\section{\label{sec:Intro}Introduction}

Antiferromagnetic (AF) nanoparticles have rich magnetic behavior
that can be quite different from their bulk counterparts. This
behavior is often termed ``anomalous'' and ``unexpected'', and
includes enhanced magnetic moment and
coercivity,\cite{Berkowitz_jmmmAF} exchange
bias,\cite{Berkowitz_prb,Berkowitz_jmmmAF} increase of magnetic
moment with
temperature,\cite{Seehra_prb_neutrons,NJOS_jmmm1,Morup_thermo} and
decrease of AF susceptibility ($\chi_{AF}$) with temperature below
the order temperature $T_N$ and its enhancement compared to bulk
\cite{Berkowitz_prb,Seehra_prb_neutrons,Seehra_ferrihydrite_doped,Gilles_EPJB,Gilles_JMMM}.
This last issue is the subject of the present report.

The enhancement of $\chi_{AF}$ below $T_N$ in nanoparticles
compared to bulk was predicted by N\'{e}el,\cite{Neel_af3} and
estimated to decrease with temperature \cite{Neel_af4, Neel_af5}.
The extra susceptibility ($\chi_a$) is a finite size effect termed
superantiferromagnetism.
In a simple picture, superantiferromagnetism arises in particles
in which the AF easy axis is perpendicular to the external field,
where surface spins rotate more in the field direction than inner
ones since they have less neighbors. %%
This corresponds to a progressive rotation of the AF easy axis
from surface to surface across the particle, in particles with
even number of ferromagnetic spin planes. %%
N\'{e}el also highlighted the first difficulty in finding
experimental evidence of superantiferromagnetism: the need for
magnetic particles with small sizes and controlled size
distribution \cite{Neel_af5}. Other difficulties became apparent
later and are related to the fact that AF nanoparticles have an
uncompensated magnetic moment $\mu_{un}$ superposed to
$\chi_{AF}$. $\mu_{un}$ hinders the determination of
$\chi_{AF}(T)$ based on low field and high field susceptibility
measurements. In the case of low field measurements, the
difficulty arises since $\mu_{un}$ has an important Curie-like
contribution that is not straightforward to model, due to the fact
that the temperature dependence of $\mu_{un}$ is not yet clear
\cite{Morup_thermo,NJOS_prl,Morup_answer}. In the case of high
field measurements, the influence of $\mu_{un}$ is more subtle and
is related to the non-saturation of the magnetization associated
to $\mu_{un}$ ($M_{\mu}$) at the normally used high fields ($
5\times 10^4$ Oe) and temperatures of interest. Again, the absence
of a reliable model of the field dependence of $M_{\mu}$, nor even
of its approach to saturation, makes the separation between the
contribution of $\chi_{AF}$ and $\mu_{un}$ to the total
magnetization (and the subsequent determination of $\chi_{AF}(T)$)
quite difficult.

Despite all these questions, some steps were made towards the
determination of $\chi_{AF}(T)$.
In a first approach, $M_{\mu}(H)$ was modelled with a Langevin
law,\cite{ferritin_JMMM, Berkowitz_prb} which enabled the first
report on $\chi_{AF}(T)$ \cite{Berkowitz_prb}. In
Ref.\cite{Berkowitz_prb} and in following
ones,\cite{Seehra_prb_neutrons,Seehra_ferrihydrite_doped}
$\chi_{AF}(T)$ was found to decrease with temperature, and this
decrease was associated to superantiferromagnetism
\cite{Berkowitz_prb}. Evidence of superantiferromagnetism based on
a description of magnetization taken at 2 K up to $30\times10^{4}$
Oe was later reported in Ref.\cite{Gilles_JMMM}. The model used
for $M_{\mu}(H)$ was further refined by the use of a distribution
and an Ising-like function that takes into account the coupling
between $\mu_{un}$ and the AF moments \cite{Gilles_EPJB,
Gilles_JMMM}. Yet, these improvements did not change the observed
decrease of $\chi_{AF}(T)$. A method for the separation between
the $\chi_{AF}(T)$ and $\mu_{un}$ components in the magnetization
without the need of a model was also proposed;\cite{NJOS_prb}
however, this method does not take into account anisotropy
effects, which are relevant in antiferromagnetic nanoparticles, as
highlighted in Ref.\cite{Morup_jmmm_Aniso}. It also became clear
in Ref. \cite{Morup_jmmm_Aniso} that a spurious contribution to
$\chi_{AF}(T)$ arises when modelling $M_{\mu}(H)$ without
considering anisotropy. This spurious contribution decreases with
increasing temperature towards zero as anisotropy energy becomes
small compared to $k_BT$ and $\mu_{un}H$.

Given this scenario, a better insight on $\chi_{AF}(T)$ depends on
measurements of the susceptibility at fields higher than those
used up to now. With this aim, we present measurements taken up to
different maximum fields and different techniques of measuring
magnetization in ferritin, a model system for nanoparticles with
AF interactions where many of the above cited studies where
performed \cite{Berkowitz_prb, ferritin_JMMM, Gilles_EPJB,
Gilles_JMMM, NJOS_prb}. We study the dependence of the derived
$\chi_{AF}(T)$ on the field at which it is considered and we
discuss its origin. We compare $\chi_{AF}(T)$ estimated at the
highest measured fields to that estimated from mean field and from
mean field considering superantiferromagnetism.
%%%%%
%%Novo
%%%%%%
We also discuss the absence of a spin-flop transition in ferritin
up to $50\times 10^4$ Oe in terms of the random local anisotropy
model.

\section{Experimental}

Ferritin consists of a hollow spherical shell composed of 24
protein subunits surrounding a ferrihydrite-like core. The
diameter of the cavity is of the order of 7-8 nm and average size
of the core of horse spleen ferritin is 5 nm \cite{Mann_livro}.
Horse spleen ferritin samples used in these experiments were
obtained from Sigma Chemical Company and prepared in powder
samples by evaporation of the solvent at room temperature. The
iron content (14.25 \% in weight) was determined by inductively
coupled plasma spectrometry.
Ac susceptibility was determined as a function of temperature
after cooling in the absence of field, at selected frequencies
(33, 476 and 1379 Hz) and a field amplitude of 4 Oe, using a
MPMS-XL Quantum Design system.
Magnetization was determined as a function of field i) up to
$9\times10^{4}$ Oe at different temperatures using a PPMS system
(Quantum Design) with a vibrating sample magnetometer (VSM)
option, ii) up to $29/30\times10^{4}$ Oe at different temperatures
using an extraction magnetometer in a Bitter magnet (HFML
facility, Nijmegen), and iii) up to $50\times10^{4}$ Oe at 4.2 K
using pick up coils and a pulsed field (LNCMP facility, Toulouse).
Magnetization curves obtained in ii) and iii) were scaled with
respect to those obtained in i). Concerning curves obtained in
ii), scaling constitutes a small correction ($< 5 \%$) and all
analysis and conclusions here presented do not depend on this
scaling.

\section{Results and discussion}

\subsection{Magnetization and high field susceptibility}

\begin{figure}[htb!]
\begin{center}\includegraphics[width=0.9\columnwidth]{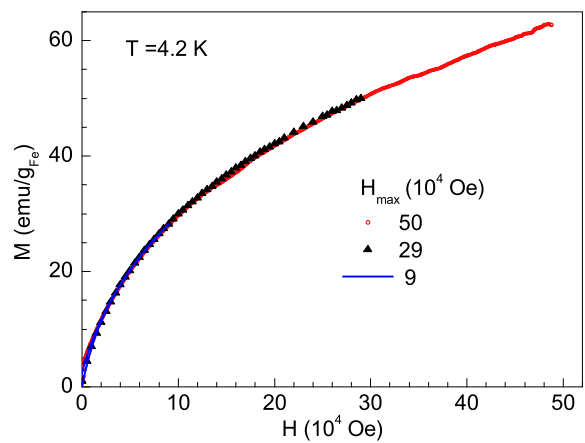}
\end{center}
\caption{\label{FigMH4K} (color online) Magnetization curves of
ferritin at 4.2 K taken up to $50\times10^4$ Oe (pulsed fields),
$29\times10^4$ Oe (static fields, extraction magnetometer), and
$9\times10^4$ Oe (static fields, VSM).}
\end{figure}

\begin{figure}[htb!]
\begin{center}\includegraphics[width=0.9\columnwidth]{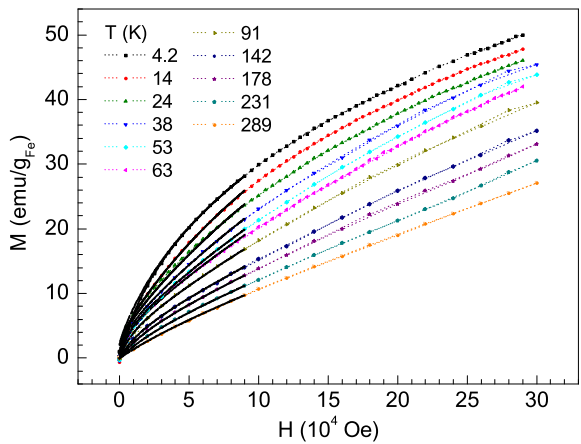}
\end{center}
\caption{\label{FigMH} (color online) Magnetization curves of
ferritin at selected temperatures, taken up to $29/30 \times10^4$
Oe (points) and taken up to $9\times10^4$ Oe (lines).}
\end{figure}

\begin{figure}[htb!]
\begin{center}\includegraphics[width=0.9\columnwidth]{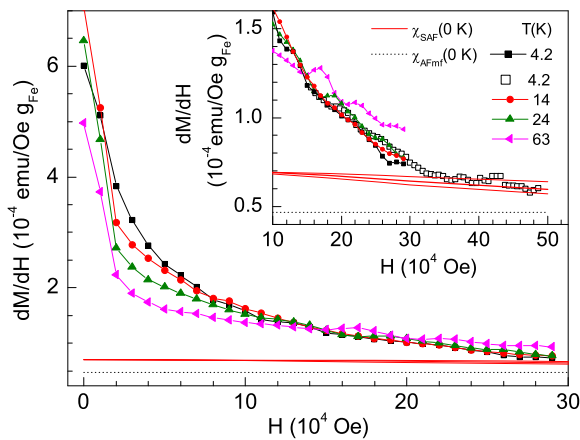}
\end{center}
\caption{\label{FigDer} (color online) Derivative of the
magnetization curves taken up to $29\times10^4$ Oe (static fields,
extraction magnetometer) as a function of field for selected
temperatures. Dotted line shows $\chi_{AF}$ expected from bulk
mean field at 0 K (termed $\chi_{AFmf}$) and continuous lines
represent the antiferromagnetic susceptibility considering
superantiferromagnetism ($\chi_{SAF}$) also at 0 K, for $2N$=10,
15 and 20. Inset shows zoom over the high field region, including
$dM/dH$ values obtained up to $50\times10^4$ Oe at 4.2 K (pulsed
fields).}
\end{figure}

The scaled magnetization curves taken up to $9\times10^{4}$,
$29/30\times10^{4}$, and $50\times10^{4}$ Oe at 4.2 K are shown in
Fig. \ref{FigMH4K} (in emu per grams of iron).
%%%%%%%%%%%%%%%%%%%%%%%%%
%%% INICIO novo, 2a ronda %%%%%%
%%%%%%%%%%%%%%%%%%%%%%%%%
The magnetization curve and its derivative (see Fig. \ref{FigDer})
have no signs of a spin-flop transition.
%%%%%%%%%%%%%%%%%%%%%%%%%
%%% FIM novo, 2a ronda %%%%%%
%%%%%%%%%%%%%%%%%%%%%%%%%
%%
%%
On the contrary, after the initial fast saturation that occurs up
to $\sim 6\times10^{4}$ Oe, the magnetization undergoes a slow
approach to saturation. %%
Clearly, $\mu_{un}$ is not yet saturated (i. e., magnetization is
not yet linear with field) at fields of the order of those often
used to estimate $\chi_{AF}$ ($5\times10^{4}$ Oe).
%%%%%%%%%%%%%%%%%%%%%%%%%
%%% INICIO novo, 2a ronda %%%%%%
%%%%%%%%%%%%%%%%%%%%%%%%%
%%%
Both the slow approach to saturation and the absence of a
spin-flop are in accordance to the previous high field
measurements performed in horse spleen ferritin at low temperature
(at 2 K and up to $30\times10^{4}$ Oe \cite{Gilles_JMMM} and at
1.52 K and up to $55\times10^{4}$ Oe \cite{ferritin_high_field}).
The slow approach to saturation is also observed in $M(H)$ curves
obtained at different temperatures (Fig. \ref{FigMH}). %%
%%%%%%%%%%%%%%%%%%%%%%%%%
%%% FIM novo, 2a ronda %%%%%%
%%%%%%%%%%%%%%%%%%%%%%%%%
%%%
%%
%%%
However, as temperature increases, the magnetization approaches a
linear regime at lower fields, i. e., at higher temperatures, the
derivative of magnetization with respect to the field $dM/dH$
approaches a nearly constant value for lower fields (Fig.
\ref{FigDer}).

\begin{figure}[htb!]
\begin{center}\includegraphics[width=0.9\columnwidth]{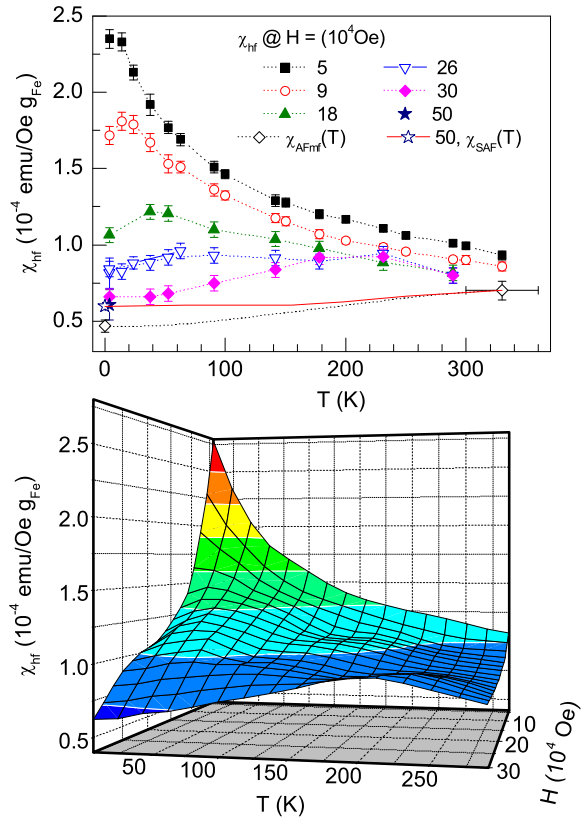}%FigChiAF.eps}
\end{center}
\caption{\label{FigChiAF} (color online) Above: high field
susceptibility $\chi_{hf}$ as a function of temperature at
selected field values, antiferromagnetic susceptibility estimated
from mean field ($\chi_{AFmf}(T)$, expected for bulk materials)
and antiferromagnetic susceptibility estimated from mean field
considering the contribution of superantiferromagnetism
($\chi_{SAF}(T)$, expected for nanoparticles) at $50\times10^4$ Oe
. Below: $\chi_{hf}$ as a function of temperature and field
obtained from magnetization curves taken up to $30\times10^4$ Oe.}
\end{figure}

With the values of $dM/dH$ it is possible to study the different
evolutions of $\chi_{AF}$ with temperature, when $\chi_{AF}$ is
estimated at different field values. In order to distinguish
between $dM/dH$ taken at a given field and the real $\chi_{AF}$
obtained for complete $\mu_{un}$ saturation, we term the
susceptibilities obtained at different (high) fields as high field
susceptibility $\chi_{hf}=dM/dH$. In Fig. \ref{FigChiAF} it is
possible to observe that $\chi_{hf}$ decreases with temperature
when taken at $5\times10^4$ Oe, in accordance with previous
results \cite{Berkowitz_prb, Gilles_EPJB, Gilles_JMMM, NJOS_prb}.
When taken at $9\times10^4$ Oe, $\chi_{hf}$ has a non-monotonic
behavior, increasing and then decreasing with temperature. For
$H=30\times10^4$ Oe, $\chi_{hf}$ is reduced about 3 times compared
to the values at  $5\times10^4$ Oe and increases with temperature
from 4.2 to about 180 K. An even lower value of $\chi_{hf}$ is
obtained at 4.2 K and $50\times10^4$ Oe. This clearly shows that
the temperature dependence of the estimated $\chi_{AF}$ depends on
the field at which it is considered, with the trend to increase
with temperature being more evident as the field increases. The
``anomalous'' behavior of $\chi_{AF}$ decreasing with temperature
for $T<T_N$ almost vanishes when $\chi_{AF}$ considered at
sufficiently high fields.
This is in agreement with a recently published Monte Carlo
simulation of AF nanoparticles with an even number of planes,
where the simulated susceptibility increases with temperature
\cite{MC_thermo}. %%

\subsection{The absence of a spin-flop transition}\label{Sec:SpinFlop}

As previously reported,\cite{ferritin_high_field} there is no
evidence of a spin-flop transition up to $50\times10^{4}$ Oe in
ferritin. In Ref. \cite{ferritin_high_field}, a spin-flop
transition was more likely to occur than in the present case,
since the temperature was lower and the field higher. From
mean-field theory, at 0 K, the spin-flop field is
%$H_{sf}=(2K/(\chi_\perp-\chi\parallel))^{1/2}$ (where $K$ is the
%anisotropy constant,
$H_{sf}=(2 H_E H_K)^{1/2}$ (where $H_E$ is the exchange field and
$H_K$ the anisotropy field), which in ferritin is about
$10\times10^{4}$ Oe accordingly to estimations of $H_E$ and $H_K$
of Ref.\cite{Gilles_JMMM}. %%%%%
%%%%
As discussed in Ref.\cite{ferritin_high_field}, the experimental
evidence of the absence of a spin-flop up to $50\times10^{4}$ Oe
implies an enhancement of $H_K$ and/or $H_E$ compared to that
expected. The absence of a spin-flop in this field range may also
be due to the relatively large uncompensated moment of ferritin,
as highlighted in Ref.\cite{Morup_review_AF}, and both reasons are
most probably related. %%%
%%%
In general $H_K$ is estimated from the anisotropy constant $K$
associated to the blocking process and from a saturation (or
sublattice) magnetization as $H_K=K/M_0$. $K$ is often estimated
by dividing the activation energy $E$ by the average nanoparticle
volume since in nanoparticles with intraparticle ferromagnetic
interactions $E=KV$. Since $E=255$ K (see
Sec.\ref{Sec:roleAnisLowMoments}) and the average ferritin core
has $N=2500$ Fe ions \cite{Mann_livro}, the average anisotropy
constant per Fe ion of the average core is $K=1.4\times10^{-17}$
erg/Fe$_{\textrm{ion}}$. Taking the sublattice magnetization
$m_0=3.2$ $\mu_B/$Fe$_{\textrm{ion}}$ (see Sec. \ref{Sec:bulkAF})
the anisotropy field is $H_K=K/m_0=470$ Oe and so
$H_{sf}=7\times10^{4}$ Oe (see estimation for $H_E$ in
 Sec. \ref{Sec:bulkAF}) in accordance with previous estimations \cite{Gilles_JMMM,
 Morup_review_AF}.
%%%%%
However, $E=KV$ does not hold in AF nanoparticles, where in
general $E\propto V^p$, with $p<1$. In fact, it was recently shown
that in ferrihydrite the energy barrier is proportional to the
square root of the total volume (i. e. $p=1/2$), corresponding to
a random distribution of energy barriers and probably of
uncompensated ions \cite{NJO_dist}. This means that in each
particle, the effective value of $E$ is given by the fluctuation
of the local anisotropy energy, such that the local anisotropy
constant $K'$ is higher than the average value calculated by
$K=E/V$, being higher by a factor of $N^{1/2}$
where $N$ is the number of Fe ions. %%%
%%%%
In other words, the energy of a nanoparticle with $N$ Fe ions and
the same local anisotropy energy of ferritin but without a random
distribution of anisotropy barriers would be
\begin{equation}
\label{Eq:Klocal} E'=K'V=E N^{1/2}.
\end{equation}
$K'$ is the barrier that each moment experiences and so we can
associated it to the spin-flop process. By doing so, we can define
a local anisotropy field $H'_{K}=H_K N^{1/2}$ and a local
spin-flop field $H'_{sf}=(2 H_E H'_K)^{1/2}$ whose estimated
value, $46 \times10^{4}$ Oe, is close to the maximum field here
used.
%%%
Therefore, the experimental absence of a spin-flop in the field
range here used can be, at least, partially explained in the frame
of the mean field
considering that $E\propto V^{1/2}$. %%%%
 We also emphasize that while the blocking is primarily probing the
 anisotropy energy experienced by the uncompensated moments in their process
of crossing the energy barrier between easy directions, the
flopping process is primarily associated to the anisotropy
experienced by the AF coupled moments, and the anisotropy field
associated to AF moments can be significantly different from that
of the uncompensated moments.%%%
%%%%

\subsection{Bulk antiferromagnetic and superantiferromagnetic
susceptibilities}\label{Sec:bulkAF}

As one might expect, the study of the enhancement of $\chi_{AF}$
in nanoparticles and of the temperature dependence of $\chi_{AF}$
benefits from comparing to bulk results. This is not possible for
ferritin, since ferrihydrite exists only in the form of
nanoparticles \cite{Ferrihydrite_review}. However a comparison to
mean field estimations can be made. In the mean field context, the
perpendicular AF susceptibility $\chi_\perp$ is %%
\begin{eqnarray}
\label{susPer} \chi_\perp=\frac{M_0}{H_E} \\\nonumber  H_E=\frac{3
k_B T_N}{m_0}
\end{eqnarray}
where $m_0$ and $M_0$ are the magnetic moment and magnetization of
an AF sublattice at 0 K, respectively, and $H_E$ the
inter-sublattice exchange field. At $T=T_N$, $\chi_{AF}$ estimated
from mean field $\chi_{AFmf}$ is equal to $\chi_\perp$, and at
$T=0$ K $\chi_{AFmf}=\frac{2}{3}\chi_\perp$. Eq.\ref{susPer}
disregards the anisotropy field, which is a good approximation for
estimating $\chi_{AFmf}$ of ferritin, since it is about 2 orders
of magnitude lower than $H_E$. Concerning the values to be used in
Eq.\ref{susPer}, there is a broad range of $T_N$ estimated for
ferritin and ferrihydrite (typically from 300 to 500 K), depending
on the used technique
\cite{Seehra_prb_neutrons,Schwertmann_neutrons,Berkowitz_prb,
Gilles_JMMM, NJOS_prb}. Using magnetization measurements, $T_N$ is
always obtained by extrapolation \cite{Berkowitz_prb, Gilles_JMMM,
NJOS_prb}, and is higher than that obtained as a direct result
with neutron diffraction for ferrihydrite ($330\pm30$ K in
Ref.\cite{Schwertmann_neutrons} and $\simeq 350$ K in
Ref.\cite{Seehra_prb_neutrons}). Neutron diffraction also gives an
estimation of the magnetic moment: $m$(5 K)=3.2
$\mu_B/$Fe$_{\textrm{ion}}$ \cite{Schwertmann_neutrons}. This
value is lower than that of isolated Fe ions $5\mu_B$ (previously
used in the estimation of $\chi_\perp$ \cite{Gilles_JMMM}) but is
reasonable for a compound where magnetic exchange interactions are
influenced by a high degree of structural disorder
\cite{Schwertmann_neutrons}. $m_0$ can be further obtained by
extrapolation to 0 K using the mean field temperature dependence
for the magnetic moment. Using the neutron diffraction results
$H_E=456\pm40\times10^{4}$ Oe and
$\chi_\perp=7.0\pm0.6\times10^{-5}$ emu/Oe g$_{Fe}$. $\chi_{AFmf}$
at 0 and $T_N$ thus estimated are plotted in Fig. \ref{FigChiAF}.

Based on $\chi_{AFmf}$ estimation it is also possible to further
estimate the antiferromagnetic susceptibility expected when
considering superantiferromagnetism $\chi_{SAF}$ (both temperature
and field dependent). At zero field, and considering only first
neighbor exchange, the perpendicular susceptibility of particles
with even number of ferromagnetic spin planes $2N$ is
$\chi_{\perp2N}=2\chi_\perp$, and considering $n^{th}$ neighbor
interactions would increase this estimation \cite{Neel_af5,
Gilles_JMMM}. By increasing the field, $\chi_{\perp2N}$ approaches
$\chi_\perp$, being this approach dependent on a characteristic
field given by $h=H_E/2N$. For low $h$ values the relation between
$\chi_{\perp2N}$ and $\chi_\perp$ is
\begin{equation}
\label{sus_prep_2N}
\frac{\chi_{\perp2N}}{\chi\perp}=2-\frac{4h^2}{3}
\end{equation}
For $h$ around unity, $\chi_{\perp2N}/\chi_\perp$ can be obtained
by solving an integral equation \cite{Neel_af5, Gilles_JMMM},
whose results are given in tables in Ref.\cite{Neel_af5}.
The perpendicular susceptibility of a set of nanoparticles with
half of them having $2N$ even can be written as
\begin{equation}
\label{sus_prep_SAF} \chi_{\perp
SAF}=\frac{1}{2}\chi_\perp+\frac{1}{2}(\chi_\perp+\chi_a)
\end{equation}
where the extra susceptibility $\chi_a=\chi_{\perp2N}-\chi_\perp$
is a function of $H$ and $T$ and can be expressed as
$\chi_a=k(H,T)\chi_\perp$. The susceptibility of a set of randomly
orientated nanoparticles can then be estimated as
\begin{equation}
\label{sus_SAF}
\chi_{SAF}(H,T)=\frac{2}{3}\left(\chi_\perp+\frac{1}{2}k(H,T)\chi_\perp\right)+\frac{1}{3}\chi_\parallel(T)
\end{equation}
In the frame of mean field, considering two sublattices with
negligible intralattice exchange interaction, the temperature
dependence of $\chi_\parallel(T)$ is given by \cite{LidiardAF}
\begin{equation}
\label{susParalela} \chi_\parallel(T)=\frac{N g^2 \mu_{B}^{2}S^2
B_{S}^{'}(y)} {k_B(T+3T_N S(S+1)^{-1}B_{S}^{'}(y))}
\end{equation}
with
\begin{eqnarray}
\label{susTotal_expli} y=\frac{3 T_N M}{(S+1)T}
\\\nonumber M=S B_{S}(y)
\end{eqnarray}
 Considering $S$ of Fe$^{3+}$,  $\chi_{AFmf}(T)$ can be readily obtained (Fig. \ref{FigChiAF},
 upper panel). $\chi_{SAF}(H,T)$ can be further calculated by estimating
 $k(H,T)$, which depends on $h$ (i. e. on $2N$ and $H_E$) and on $T/T_N$. Using $2N=15\pm5$,\cite{Gilles_JMMM} and calculating $k(H,T)$ based on
 Eq.\ref{sus_prep_2N} for $h<0.3$, and on the tables presented in
 \cite{Neel_af5} for $h>0.3$ and $T/T_N>0.2$, one can estimate
 $\chi_{SAF}(H,T)$, plotted as a function of field for $T=0$K in
 Fig. \ref{FigDer} (using the average value of $2N$ and its upper and lower limits)
 and as a function of temperature for $H=50\times10^4$
 Oe in Fig. \ref{FigChiAF} upper panel. It is clear from Fig. \ref{FigDer} that up
 to fields of the order of $25\times10^4$ $dM/dH$ is higher than
 that expected from  $\chi_{SAF}$ and $\chi_{AFmf}$, having thus contribution from
 mechanisms other than bulk antiferromagnetism and
 superantiferromagnetism. For $H>29\times10^4$ Oe and at 4.2 K, $dM/dH$ approaches
 $\chi_{SAF}$ and $\chi_{AFmf}$, and bulk antiferromagnetism and
 superantiferromagnetism are the relevant contributions for
 $dM/dH$ (Fig. \ref{FigDer}, inset).
At $50\times10^4$ Oe and 4.2 K, $dM/dH$ is of the order of
$\chi_{SAF}$ and $\chi_{AFmf}$, being closer to $\chi_{SAF}$ than
$\chi_{AFmf}$. Therefore, considering that at this temperature and
field $\mu_{un}$ is already saturated the AF susceptibility has a
contribution from superantiferromagnetism.
%
%%%%%
%%%%%
%
Concerning the temperature dependence of the susceptibilities,
(Fig. \ref{FigChiAF}, upper panel), it is clear that for lower and
higher temperatures $\chi_{hf}$ at $30\times10^4$ Oe is close to
$\chi_{SAF}$, while it deviates in the intermediate temperature
region, this deviation being higher than the difference between
$\chi_{SAF}$ and $\chi_{AFmf}$. It is also noteworthy that, while
$\chi_{\perp2N}$ decreases with temperature, due to the approach
of $\chi_{\perp2N}$ to $\chi_{\perp}$ at high fields, due to
averaging particles with even and odd $2N$, averaging
$\chi_\parallel$ and $\chi_\perp$, and due to the temperature
increase of $\chi_\parallel$, $\chi_{SAF}$ at $50\times10^4$ Oe
estimated for ferritin is roughly constant up to $T=150$ K
increasing then with temperature up to $T_N$.

\subsection{The role of anisotropy and small magnetic moments}
\label{Sec:roleAnisLowMoments}

\begin{figure}[htb!]
\begin{center}\includegraphics[width=0.9\columnwidth]{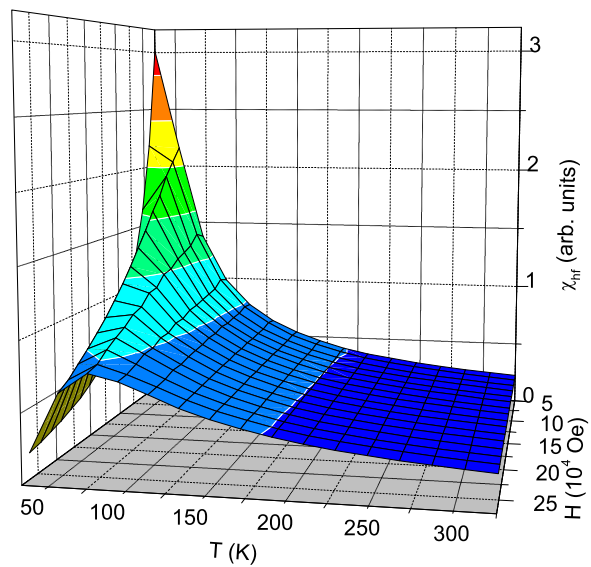}%FigChiAF.eps}
\end{center}
\caption{\label{FigSimLang} (color online) $\chi_{hf}=dM/dH$ as a
function of temperature and field obtained from magnetization
curves simulated using the Langevin function and $\mu_{un}=5$
$\mu_B$.}
\end{figure}

From the above discussion, it is clear that
superantiferromagnetism is not the most relevant mechanism
responsible for the fact that $\chi_{hf}$ is larger than
$\chi_{AF}$ for $H<28\times10^4$ neither for the decrease of
$\chi_{hf}$ with temperature below $T_N$.
It is therefore interesting to investigate the origin of this
enhancement and decrease with temperature, which is expected to be
related to the non-saturation of $\mu_{un}$. In turn, this
non-saturation is either due to small $\mu_{un}$ values or due to
the role of anisotropy in the approach to saturation of $M_{\mu}$.
The existence of such small moments, in particular paramagnetic
Fe$^{3+}$ ions, is in fact expected from relaxometry results
\cite{ferritin_relaxometry_magnet}.
These small moments may in fact be related to the non-monotonic
behavior observed in the temperature dependence of $\chi_{hf}$.
Simple calculations of an hypothetical contribution of small
moments to $\chi_{hf}$ based on $dM/dH$ with $M(H)$ being given by
the Langevin law show that $dM/dH$ increases and then decreases
with temperature at a given field, with the temperature at which
that maximum occurs increasing with the field (Fig.
\ref{FigSimLang}). This behavior is qualitatively similar to that
observed in Fig. \ref{FigChiAF}.
Concerning the role of anisotropy, although for fields of the
order of $10\times10^4$ Oe , for the typical average $\mu_{un}$ of
ferritin ($\sim 100$ $\mu_B$) and $T\sim5$ K, $M_{\mu}$ would be
close to saturation accordingly to the Langevin law, anisotropy
retards saturation to higher fields so that the typical average
$\mu_{un}$ still gives an important contribution to $\chi_{hf}$ in
the above mentioned conditions. As temperature increases, the
relevance of anisotropy energy decreases compared to $k_BT$,
leading to a decrease of the contribution of $M_{\mu}$ to
$\chi_{hf}$. Considering only the $M_{\mu}$ component and two
temperatures $T_2>T_1$, there is a cross-over field below which
$\chi_{hf}(T_2)<\chi_{hf}(T_1)$, so that $M_{\mu}$ gives a
spurious contribution to the total $\chi_{hf}$ that decreases with
temperature. This cross-over field increases in comparison to the
Langevin law when uniaxial anisotropy is considered. When surface
anisotropy is further taken into account, the increase of this
cross-over field is quite dramatic: accordingly to simulations
shown in Ref.\cite{Kachkachi_efect_anis_MH}, the cross-over field
is of the order of $8\times10^4$ Oe for Co nanoparticles (515
spins) and temperatures between $\sim0.5$ and $\sim10$ K.

%%%%%%%%%%%%%%%%%%%%%%%%%
%%% INICIO novo, 2a ronda %%%%%%
%%%%%%%%%%%%%%%%%%%%%%%%%

%
 \begin{figure}[htb!]
\begin{center}\includegraphics[width=0.7\columnwidth]{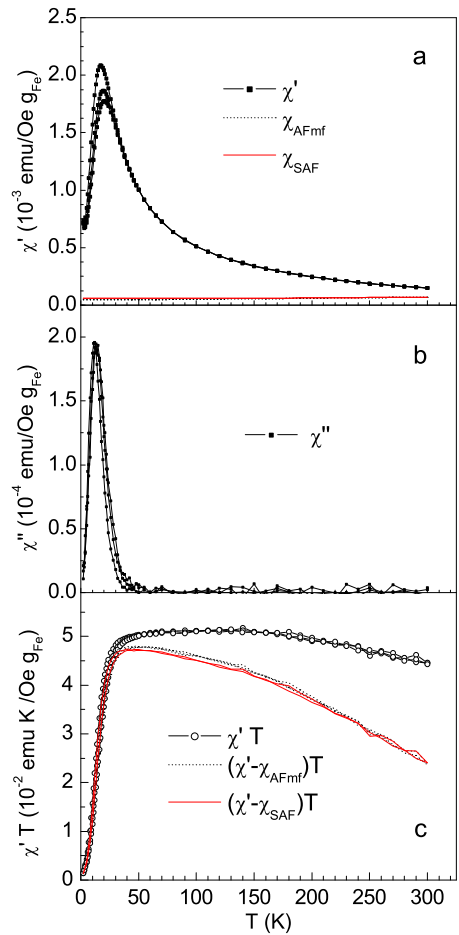}%FigChiAF.eps}
\end{center}
\caption{\label{FigSus}(color online) a. temperature dependence of
the in phase component of the ac susceptibility $\chi'$ for
different field frequencies (33, 476 and 1379 Hz), $\chi_{SAF}(T)$
estimated considering superantiferromagnetism and $\chi_{AFmf}(T)$
estimated from mean field; b. temperature dependence of the out of
phase component of the ac susceptibility $\chi''$ and c.
temperature dependence of the $\chi'T$ product and temperature
dependence of the product of temperature and susceptibility
associated to $\mu_{un}$ determined as $(\chi'-\chi_{AFmf}(T))T$
and $(\chi'-\chi_{SAF}(T))T$.}
\end{figure}

\begin{figure}[htb!]
\begin{center}\includegraphics[width=0.9\columnwidth]{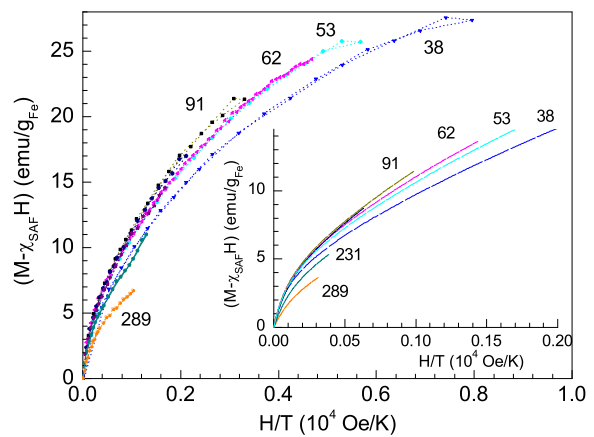}%FigChiAF.eps}
\end{center}
\caption{\label{FigMsat}(color online) Saturation component of the
magnetization, obtained after subtracting $\chi_{SAF}(T)H$ to the
total magnetization, in a $H/T$ scale. Inset shows a zoom over a
lower field region, concerning data obtained with VSM.}
\end{figure}

The existence of a relevant anisotropy contribution to $M(H)$
curves can be qualitatively evaluated combining information from
$M_{\mu}$ in a $H/T$ scale and ac susceptibility, since anisotropy
does not affect the equilibrium linear susceptibility ($\chi'$
above blocking), affecting
$M_{\mu}$ at intermediate fields whenever relevant. %%%
%%%%
%%%%
The in phase ($\chi'$) and out of phase ($\chi''$) components of
the ac susceptibility (Fig. \ref{FigSus}) show characteristic
features of ferritin superparamagnetic nanoparticles, namely
frequency dependence below temperatures of the order of 40 K and a
frequency dependent maximum at around 20 K
\cite{FLuis_prb_ferritin}. %%%
From the frequency dependence of the maximum of $\chi''$ it is
possible to estimate an energy barrier associated to the blocking
process as 255 K.
%%%%
%%%
%%%%
%%%%
%%%%%
The antiferromagnetic susceptibility $\chi_{SAF}(T)$ and
$\chi_{AFmf}(T)$ can be subtracted to $\chi'$, in order to study
the temperature dependence of $\mu_{un}$ based on a susceptibility
temperature product plot, since interparticle interactions are
negligible \cite{FLuis_prb_ferritin}.
%%%%%%
%%%%%
%%%%%%% novo
 %%%
$(\chi'-\chi_{SAF}(T))T$ corresponds also to the slope of
$M_{\mu}=M-\chi_{SAF}(T)H$ in a $H/T$ scale at $H=0$. %%
Both $(\chi'-\chi_{SAF}(T))T$ and $(\chi'-\chi_{AFmf}(T))T$
increase with temperature up to $~40$ K, the temperature at which
$\chi''$ becomes zero, corresponding to an increase of the average
$\mu_{un}$ due to the unblocking process (Fig. \ref{FigSus}c). For
$T>40$ K, $(\chi'-\chi_{SAF}(T))T$ and $(\chi'-\chi_{AFmf}(T))T$
decrease with temperature due to the decrease of the sublattice
magnetization when approaching $T_N$,\cite{Gilles_JMMM,NJOS_prb}
being this decrease more pronounced for $T\gtrsim 90$ K. %%%%
Due to the decrease of $(\chi'-\chi_{SAF}(T))T$ with temperature
for $T>40$ K (above blocking), $M_{\mu}$ is not expected to scale
in a $H/T$ plot, being expected lower values for $M_{\mu}$ in the
curves taken at higher temperatures in all the $H/T$ range. In
Fig. \ref{FigMsat} it is clear that above blocking
$M_{\mu}=M-\chi_{SAF}(T)H$ does not scale in $H/T$ for $H/T>100$
Oe/K. In particular, in the $38<T<91$ K range and $H/T>100$ Oe/K,
$M_{\mu}$ (and the slope of $M_{\mu}$) is higher in the curves
taken at higher temperatures (Fig. \ref{FigMsat}, inset), unlike
that expected from the slightly decrease of
$(\chi'-\chi_{SAF}(T))T$. For $T>91$ K, $M_{\mu}$ in a $H/T$ scale
is always lower in the curves taken at higher temperatures, as
expected from the
decrease of $(\chi'-\chi_{SAF}(T))T$. %%%%
%%%%
%%%%
%
In other words, the non-scaling of $M_{\mu}$ in the $38<T<91$ K
temperature range and intermediate $H/T$ values where $M_{\mu}$
has higher values for curves taken at higher temperatures cannot
be explained from the behavior of $\mu_{un}(T)$. Since
$\mu_{un}(T)$ cannot account for the behavior of $M_{\mu}(H/T)$,
since interparticle interactions in ferritin are negligible
\cite{FLuis_prb_ferritin} and a distribution of uncompensated
moments for it self does not produce a non-scaling of $M_{\mu}$
(the sum of functions of $H/T$ is also a function of $H/T$), the
only reason left for the behavior of $M_{\mu}(H/T)$ in the
$38<T<91$ K temperature range is anisotropy. In fact,
%%
%%the existence of a relevant anisotropy contribution in $M_{\mu}$
%%explains the increase of $M_{\mu}(H/T)$ for higher temperatures in
%%the $38<T<91$ K range and intermediate $H/T$ values when
%%$\mu_{un}$ does not increase with temperature, since anisotropy
%%does not affect $M_{\mu}$ in the low
%%susceptibility range. %%%
the increase of $M_{\mu}(H/T)$ for curves taken at higher
temperatures and for a given $H/T$ value in a intermediate range
and scaling (or decrease) in the low $H/T$ range is, in fact, a
fingerprint of anisotropy, as found for instance in Co
nanoparticles,\cite{Respaud_Co_jap} and in simulations
\cite{Kachkachi_efect_anis_MH,Morup_jmmm_Aniso}.
%%%FIM NOVO
%%%%
Therefore anisotropy has a relevant contribution to the $M(H)$
curves of ferritin, being one of the causes to the non-saturation
of $\mu_{un}$ at the applied fields normally used.

\section{Conclusions}

We show that the derived $\chi_{AF}(T)$ depends critically on the
maximum field at which it is determined. %%%
%%%%%
%%%%
When it is determined at fields of the order of $5\times10^{4}$
Oe, $\chi_{AF}$ decreases with temperature, similarly to earlier
studies \cite{Berkowitz_prb,Gilles_EPJB, Gilles_JMMM}. This
behavior is related to the influence of anisotropy in the approach
to saturation of $\mu_{un}$ and probably due to the existence of
small magnetic moments, that leads to the non saturation of
$M_{\mu}$ at fields of the order of $5\times10^{4}$.
%%%%
%%%%
%%%%
 On the contrary, when $\chi_{AF}$ is determined as $dM/dH$ at $30\times10^{4}$ Oe,
 it increases with temperature for $4.2<T<180$ K (i. e. below $T_N$)
as in bulk AF.  %%%%
%%%%%
%%%%
At fields of the order of $50\times10^{4}$ Oe and at 4.2 K,
$\chi_{AF}$ determined as $dM/dH$ is in good agreement to
$\chi_{AF}$ estimated from mean field considering the effect of
superantiferromagnetism, and of the order of $\chi_{AF}$ estimated
from mean field.

%%%This increase is in accordance to recently published Monte Carlo
%%%simulations \cite{MC_thermo}.

\begin{acknowledgments}
The authors acknowledge A. Urtizberea for helping with the high
field measurements and for fruitful discussions, and E. Lythgoe
for critical reading the manuscript. Part of this work has been
supported by EuroMagNET under the EU contract RII3-CT-2004-506239.
The financial support from FCT, POCTI/CTM/46780/02, bilateral
projects GRICES-CSIC, and Acci\'{o}n Integrada Luso-Espa\~{n}ola
E-105/04 are gratefully recognized. The work in Zaragoza has been
supported by the research grants MAT2007-61621 and
CONSOLIDER-INGENIO 2010 Programme, grant CSD2007/0010 from the
Ministry of Education. N. J. O. S. acknowledges CSIC for a I3P
contract.
\end{acknowledgments}

%\newpage %Just because of unusual number of tables stacked at end
\bibliography{bib_NJOSilva}% Produces the bibliography via BibTeX.

\end{document}